\documentclass[
 aip,
 amsmath,amssymb,
 preprint,
]{revtex4-2}

\usepackage[utf8]{inputenc}
\usepackage{graphicx}
\usepackage[dvipsnames]{xcolor}
\usepackage{braket}
\usepackage{algorithm}
\usepackage{algpseudocode} 

\newcommand{\vect}[1]{\boldsymbol{#1}}
\newcommand{\matv}[1]{\mathbf{#1}}
\newcommand{\mat}[1]{\text{#1}}

\newcommand{\bmat}[1]{\mathbf{#1}}
\newcommand{\bmatv}[1]{\overrightarrow{\mathbf{{#1}}}}

\begin{document}

\title{Lagrangian Formulation of Nuclear-Electronic Orbital Ehrenfest Dynamics with Real-time TDDFT for Extended Periodic Systems}

\author{Jianhang Xu}
\author{Ruiyi Zhou}
\affiliation{Department of Chemistry, University of North Carolina at Chapel Hill, Chapel Hill, North Carolina 27599, USA}
\author{Tao E. Li}
\affiliation{Department of Chemistry, Yale University, New Haven, Connecticut 06520, USA}
\affiliation{Department of Physics and Astronomy, University of Delaware, Newark, Delaware 19716, USA}
\author{Sharon Hammes-Schiffer}
\email{shs566@princeton.edu}
\affiliation{Department of Chemistry, Yale University, New Haven, Connecticut 06520, USA}
\affiliation{Department of Chemistry, Princeton University, Princeton, New Jersey  08544, USA}
\author{Yosuke Kanai}
\email{ykanai@unc.edu}
\affiliation{Department of Chemistry, University of North Carolina at Chapel Hill, Chapel Hill, North Carolina 27599, USA}
\affiliation{Department of Physics and Astronomy, University of North Carolina at Chapel Hill, Chapel Hill, North Carolina 27599, USA}

\date{\today}

\begin{abstract}
We present a Lagrangian-based implementation of Ehrenfest dynamics with nuclear-electronic orbital (NEO) theory and real-time time-dependent density functional theory (RT-TDDFT) for extended periodic systems.
In addition to a quantum dynamical treatment of electrons and selected protons, this approach allows for the classical movement of all other nuclei to be taken into account in simulations of condensed matter systems.
Furthermore, we introduce a Lagrangian formulation for the traveling proton basis approach and propose new schemes to enhance its application for extended periodic systems. 
Validation and proof-of-principle applications are performed on electronically excited proton transfer in the o-hydroxybenzaldehyde molecule with explicit solvating water molecules. 
These simulations demonstrate the importance of solvation dynamics and a quantum treatment of transferring protons. 
This work broadens the applicability of the NEO Ehrenfest dynamics approach for studying complex heterogeneous systems in the condensed phase. 
\end{abstract}

\maketitle

\section{Introduction}

\par Nuclear quantum mechanical effects, such as zero-point energy and hydrogen tunneling, are central to various physical, chemical, and biological processes.
The coupled quantum dynamics of electrons and protons are also essential for many of these processes, such as proton-coupled electron transfer (PCET).\cite{li2006ultrafast,huynh_proton-coupled_2007,hammes2010theory,weinberg_proton-coupled_2012,chen2013chemical,hammes-schiffer_proton-coupled_2015}
Transcending the typical Born-Oppenheimer approximation in describing the nuclear quantum effects represents a significant challenge in first-principles calculations.
The necessity to incorporate heterogeneous environments for such coupled quantum dynamical processes, for example, in the context of solar-fuel research,\cite{dempsey2021vision} adds another level of complexity to this challenge.

\par The nuclear-electronic orbital
(NEO)\cite{webb_multiconfigurational_2002,pavosevic_multicomponent_2020,xu_full-quantum_2020,hammes-schiffer_nuclearelectronic_2021} approach is a highly successful multicomponent formalism that treats electrons and selected nuclei such as protons quantum mechanically on equal footing, while the remaining nuclei are typically treated classically.
Many electronic structure methods, such as Hartree-Fock,\cite{webb_multiconfigurational_2002} coupled-cluster methods,\cite{pavosevic2018multicomponent} multireference methods,\cite{webb_multiconfigurational_2002,fajen2021multicomponent} and density functional theory (DFT)\cite{pak_density_2007,chakraborty_development_2008} have been adapted into the NEO framework for studying both ground and excited states properties.
Furthermore, the coupled quantum dynamics of electrons and protons can be modeled by combining the NEO formalism with explicit dynamic approaches, such as real-time time-dependent DFT (RT-TDDFT)\cite{zhao2020real,wildman2022solvated,li2023electronic} and Ehrenfest dynamics.\cite{zhao2020nuclear,zhao2021excited,chow2023nuclear} 
The RT-NEO-TDDFT approach propagates the electronic and protonic densities numerically, including all nonadiabatic effects between the electrons and protons. 
This approach can be combined with the NEO Ehrenfest dynamics approach to allow other nuclei to move classically in a manner that includes the nonadiabatic effects between these classical nuclei and the electronic/protonic quantum subsystem. 

\par Although the majority of prior studies with NEO focused on isolated molecular systems, 
recent work extended the NEO formalism to enable the investigation of condensed phase systems.\cite{wildman2022solvated,chow2023nuclear,xu2022nuclear,xu2023first}
For instance, the NEO-DFT and RT-NEO-TDDFT approaches have been expanded to encompass condensed phases using the polarizable continuum model.\cite{wildman2022solvated}
Additionally, explicit solvation effects have been investigated by combining NEO-DFT, RT-NEO-TDDFT, and NEO Ehrenfest dynamics with a hybrid quantum mechanical/molecular mechanical (QM/MM) methodology.\cite{chow2023nuclear1,chow2023nuclear}
In our recent work, we extended the multicomponent NEO-DFT\cite{xu2022nuclear} and RT-NEO-TDDFT\cite{xu2023first} methods to enable calculations of extended periodic systems. 
For example, we studied water at a TiO$_2$ interface, a molecule in explicit solvent, and photoinduced proton transfer in a molecule attached to a silicon surface within the NEO framework.

\par Herein, we address two of the remaining challenges for exploring chemical and biological processes within the NEO framework. 
First, we implement RT-NEO-TDDFT Ehrenfest dynamics for periodic systems, allowing all nuclei to move in a manner that incorporates nonadiabatic effects between the various subsystems. 
For this purpose, we develop a theory of periodic RT-NEO-TDDFT from the Lagrangian perspective and extend it to Ehrenfest dynamics. 
Second, we investigate different schemes for moving the basis function centers associated with the quantum protons for RT-NEO-TDDFT Ehrenfest dynamics approaches. 
Previous work\cite{zhao2020nuclear,zhao2021excited} devised a semiclassical traveling proton basis (TPB) approach, in which the proton basis function centers move classically with the other nuclei. 
We propose alternative schemes based on the NEO Lagrangian and analyze the advantages and disadvantages of each scheme.

\par The remainder of this paper is organized as follows.
In Sec. \ref{sec:method}, we provide an introduction to the multicomponent RT-NEO-TDDFT framework and adapt the approach for periodic systems using atomic-centered basis functions.
We then combine the RT-NEO-TDDFT method with the Ehrenfest dynamics approach using several different TPB schemes based on the Lagrangian. 
The numerical implementation of the new methods for periodic electronic structure calculations within the FHI-aims code \cite{blum2009ab} is presented.
In Sec. \ref{sec:result}, we validate our implementation within the FHI-aims code \cite{blum2009ab} and also compare the results among different TPB schemes for photoinduced proton transfer in o-hydroxybenzaldehyde (oHBA).
A proof-of-principle application to an extended system, namely oHBA in explicit water, is presented to showcase the capabilities of the periodic RT-NEO-TDDFT Ehrenfest method. 
In Sec. \ref{sec:conclusion}, we present conclusions and discuss future directions.

\section{Theory and Computational Methods}
\label{sec:method}
\subsection{RT-NEO-TDDFT Lagrangian} 
\label{sec:RT-NEO-TDDFT_L}

\par In order to derive the equation of motion (EOM) for all degrees of freedom by applying the variational principle \cite{kramer2005geometry} to the action, we first define the Lagrangian of our system.
Within the Kohn-Sham (KS) ansatz of multicomponent TDDFT, the NEO Lagrangian (in atomic units) for our coupled quantum electron-proton system with classical nuclei is 
\begin{equation}
\begin{aligned}
\label{eq:lagrangian}
    L^\text{NEO}(t)
    & = \int \text{d}\vect{r}^{e} \int \text{d}\mathbf{k} \sum_n {\psi_{n\mathbf{k}}^e}^{*} (\vect{r}^e,t)\left[ i\frac{\partial}{\partial t}+\frac{1}{2}\nabla_{\vect{r}^e}^{2} \right] \psi_{n\mathbf{k}}^e(\vect{r}^e,t) \\
    & \quad - \frac{1}{2}\int \text{d}\vect{r}^e \text{d}\vect{r}^{\prime e} \frac{1}{|\vect{r}^e-\vect{r}^{\prime e}|} \rho^e(\vect{r}^e,t)\rho^e(\vect{r}^{\prime e},t) -E_{XC}^e[\rho^e] \\
    & \quad + \int \text{d}\vect{r}^p \sum_n {\psi_n^{p}}^* (\vect{r}^p,t)\left[ i\frac{\partial}{\partial t}+\frac{1}{2M^p}\nabla_{\vect{r}^p}^{2} \right] \psi_n^{p}(\vect{r}^p,t) \\
    & \quad - \frac{1}{2}\int \text{d}\vect{r}^p \text{d}\vect{r}^{\prime p} \frac{1}{|\vect{r}^p-\vect{r}^{\prime p}|} \rho^p(\vect{r}^p,t)\rho^p(\vect{r}^{\prime p},t) 
    -E_{XC}^p[\rho^p]  \\
    & \quad +
    \int \text{d}\vect{r}^e \text{d}\vect{r}^p \frac{1}{|\vect{r}^e-\vect{r}^p|} \rho^e(\vect{r}^e,t)\rho^p(\vect{r}^p,t)-E_{epc}[\rho^e, \rho^p] \\
    & \quad + \sum_{I}^{N^c} \frac{1}{2}M_I \dot{\vect{R}}_{I}^2(t)
    -\sum_{I<J}^{N^c}\frac{Z_I Z_J}{|\vect{R}_{I}(t)-\vect{R}_{J}(t) | } \\
    & \quad + \int \text{d}\vect{r}^e \rho^e(\vect{r}^e,t) \sum_{I}^{N^c} 
    \frac{Z_I}{|\vect{r}^e-\vect{R}_I(t)|}
    - \int \text{d}\vect{r}^p \rho^p(\vect{r}^p,t) \sum_{I}^{N^c} 
    \frac{Z_I}{|\vect{r}^p-\vect{R}_I(t)|},
\end{aligned}
\end{equation}
where ${\psi_{n\mathbf{k}}^e}$ is the electron KS orbital with reciprocal space sampling $\mathbf{k}$ and ${\psi_{n}^p}$ is the proton KS orbital using $\Gamma$-point sampling of the Brillouin zone.\cite{xu2022nuclear}
$E^e_{XC}$, $E^p_{XC}$, and $E_{epc}$ represent the exchange-correlation (XC) energy of electrons, the XC energy of protons, and the correlation energy between an electron and a proton, respectively.
$\vect{R}_{I}(t)$ and $Z_I$ are the position coordinates along the trajectory and the charge, respectively, of classical nucleus $I$. 
The upper dot denotes the time derivative, and 
$N^c$ indicates summing over all classical atomic nuclei.
The electron and proton densities are given by the time-dependent KS orbitals by   
\begin{equation}
\begin{aligned}
    \rho^e(\vect r^e,t)&=\int \text{d}\mathbf{k} \sum_n^{N^e}\psi_{n\mathbf{k}}^{e*}(\vect r^e,t)\psi_{n\mathbf{k}}^{e}(\vect r^e,t), \\
    \rho^p(\vect r^p,t)&=\sum_n^{N^p}\psi_n^{p*}(\vect r^p,t)\psi_n^{p}(\vect r^p,t).
\end{aligned}
\end{equation}
We do not refer to the electron spin explicitly in this discussion for brevity. In our implementation,  both spin restricted and unrestricted calculations can be performed.

\par We derive the EOM for different degrees of freedom by applying the principle of least action to $A = \int_{t0}^{t1}L(t)\text{d}t$
with $L(t)=L^\text{NEO}(\vect R,\dot{\vect R}, 
    \{\psi_{n\bf{k}}^{e}\},
    \{\psi_{n}^{p}\},t)$ 
for our quantum-classical NEO system.
As discussed in Ref. \onlinecite{xu2023first}, applying the variational principle to the action with respect to electron and proton KS orbitals, 
\begin{equation}
    \frac{\delta A}{\delta \psi_{n\bf{k}}^{e*}(\vect r^e,t)}=0 \quad\text{and}\quad \frac{\delta A}{\delta \psi_{n}^{p*}(\vect r^p,t)}=0,
\end{equation}
leads to the time-dependent (TD) KS equations
\begin{align}
\label{eq:tdks_e}
      i\frac{\partial}{\partial t} \psi^e_{n\bf{k}} (\vect r^e,t) & = \hat{H}^e \psi^e_{n\bf{k}} (\vect r^e,t) = \left[-\frac{1}{2}\nabla^2 + U_{\text{eff}}^e(\vect r^e)\right] \psi^e_{n\bf{k}}(\vect r^e,t), \\
 \label{eq:tdks_p}
     i\frac{\partial}{\partial t}  \psi^p_n (\vect r^p,t) & = \hat{H}^p \psi^p_n (\vect r^p,t) = \left[-\frac{1}{2M^p}\nabla^2 + U_{\text{eff}}^p(\vect r^p)\right] \psi^p_n (\vect r^p,t).
\end{align}
Here the effective potentials are defined as
\begin{equation}
\label{equ:v_eff_e}
\begin{aligned}
    U_{\text{eff}}^e(\vect r^e) &= - V_{\text{es}}^{c}(\vect r^e) - V_{\text{es}}^{p}(\vect r^e) - V_{\text{es}}^{e}(\vect r^e) +\frac{\delta E_{\text{xc}}^{e}[\rho ^e]}{\delta \rho^e}+\frac{\delta E_{\text{epc}}[\rho^e,\rho^p]}{\delta \rho^e}, \\
    U_{\text{eff}}^p(\vect r^p) &= V_{\text{es}}^{c}(\vect r^p) + V_{\text{es}}^{p}(\vect r^p) + V_{\text{es}}^{e}(\vect r^p) + \frac{\delta E_{\text{xc}}^{p}[\rho ^p]}{\delta \rho^p}+\frac{\delta E_{\text{epc}}[\rho^e,\rho^p]}{\delta \rho^p},
\end{aligned}
\end{equation}
where the electrostatic potentials of the classical nuclei, quantum protons, and electrons are 
\begin{equation}
\begin{aligned}
V_{\text{es}}^{c}(\vect r; \vect R) &= \sum_{I}^{N^c} \frac{Z_I}{|\vect{r} -\vect{R}_I(t)|}, \\
V_{\text{es}}^{p}(\vect r; \rho^p) &= \int \text{d}\vect{r}^{\prime p} \frac{1}{|\vect{r}-\vect{r}^{\prime p}|}\rho^p(\vect{r}^{\prime p},t), \\
V_{\text{es}}^{e}(\vect r; \rho^e) &= \int \text{d}\vect{r}^{\prime e} \frac{-1}{|\vect{r}-\vect{r}^{\prime e}|}\rho^e(\vect{r}^{\prime e},t).
\end{aligned}
\end{equation}

\par In general, the time-dependent Schr\"odinger equation can be derived by applying the variational principle to the action in a few different ways. 
The Dirac-Frenkel approach\cite{raab2000dirac} is perhaps the most widely used in electronic structure theory. 
Treating the KS orbitals as complex-valued scalar fields $\psi(\textbf{r})$, the TD-KS equations are obtained as the EOM straightforwardly by using the Euler-Lagrange equation $$\frac{\delta L^\text{NEO}}{\delta \psi^{*}}-\frac{d}{dt}\frac{\delta L^\text{NEO}}{\delta \dot{\psi^{*}}}=0.$$
For the classical degrees of freedom in Ehrenfest dynamics, applying the variational principle to the action with respect to the classical atomic nuclei leads to the following Euler-Lagrange equation
\begin{equation}
\label{eq:EL_equation}
    \frac{\partial L^\text{NEO}}{\partial \vect R_I(t)} - \frac{\text{d}}{\text{d} t}\frac{\partial L^\text{NEO}}{\partial \dot{\vect R}_I(t)}=0. 
\end{equation}
Despite these compact expressions for the EOMs, the actual implementation using atom-centered basis sets, as is the case for the implementation in the FHI-aims code, 
is complicated when the atoms are allowed to move as in Ehrenfest dynamics. 

\subsection{RT-NEO-TDDFT  Ehrenfest Dynamics with Atomic-centered Basis Functions} 
\label{sec:RT-NEO-TDDFT_B}

\par In our numeric atomic-orbital (NAO)-based implementation, the time-dependent KS orbitals are expanded using a set of atomic-centered basis functions $\{\phi^e_i,\phi^p_i\}$
\begin{equation}
\label{eq:basis1}
\begin{aligned}
    \psi^e_{n\mathbf{k}} (\vect r^e,t) = & \sum_i c^e_{n\mathbf{k},i}(t)\phi^e_i(\vect r^e,\vect R^e_I),\\
    \psi^p_{n} (\vect r^p,t) = & \sum_i c^p_{n,i}(t)\phi^p_i(\vect r^p,\vect R^p_I).
\end{aligned}
\end{equation}
For isolated molecules, $\{\phi^e_i,\phi^p_i\}$ represents a basis function localized at a center $I$ with position \{$\vect R^e_I,\vect R^p_I$\}.
To describe extended systems, the basis function $\{\phi^e_i,\phi^p_i\}$ is expressed as a sum of its periodic Bloch functions
\begin{equation}
\label{eq:basis2}
\begin{aligned}
    \phi^{e}_{i}\left(\vect r^e,\vect R^e_I\right) & = \sum_{\vect N}e^{i\bf{k}\cdot \vect{T(N)}}\phi^{e}_{i}(\vect r^e-\vect R^e_I+\vect T(\vect N)), \\
    \phi^{p}_{i}\left(\vect r^p,\vect R^p_I\right) &=  \sum_{\vect N}\phi^{p}_i\left(\vect r^p-\vect R^p_I+\vect T(\vect N)\right),
\end{aligned}
\end{equation}
where $\vect N= (N_1,N_2,N_3)$ are the neighboring unit cells, $\vect T(\vect N)$ is a translation vector of cell $\vect N$, and $\vect R_{I}$ is the coordinate of the center of basis function $i$.
In this formulation in which the basis functions depend on $\textbf{R}_I$, the partial time derivative in Eq. \ref{eq:lagrangian} needs to be replaced by the total time derivative because the quantum nuclei are allowed to move, 
\begin{equation}
\begin{aligned}
    \frac{\partial}{\partial t}\psi^e_{n\mathbf{k}}(\vect r,t) \rightarrow \frac{\text{d}}{\text{d} t}\psi^e_{n\mathbf{k}}(\vect r,t)  &= \sum_i\frac{\text{d}}{\text{d} t} \left[c^e_{n\mathbf{k},i}(t)\phi^e_i(\vect r,\vect R^e_I)\right] \\
    &= \sum_i\left[\dot c^e_{n\mathbf{k},i}(t)\phi^e_i(\vect r,\vect R^e_I) + c^e_{n\mathbf{k},i}(t)\dot{\vect R^e_I} \frac{\partial \phi^e_i}{\partial \vect R^e_I}(\vect r,\vect R^e_I)\right], \\
    \frac{\partial}{\partial t}\psi^p_n(\vect r,t) \rightarrow \frac{\text{d}}{\text{d} t}\psi^p_n(\vect r,t) &= \sum_i\frac{\text{d}}{\text{d} t} \left[c^p_{n,i}(t)\phi^p_i(\vect r,\vect R^p_I)\right]\\
    &= \sum_i \left[\dot c^p_{n,i}(t)\phi^p_i(\vect r,\vect R^p_I) + c^p_{n,i}(t) \dot{\vect R^p_I} \frac{\partial \phi^p_i}{\partial \vect R^p_I}(\vect r,\vect R^p_I)\right].
\end{aligned}
\end{equation}
We use the following matrices for concise notation in the following discussions:
\begin{equation}
\label{eq:mat}
\begin{aligned}
    \mat T^x_{ij} := & \braket{\phi^x_i|-\frac{1}{2M^x}\nabla^2|\phi^x_j}, \\
    \mat H^x_{ij} := & \braket{\phi^x_i|-\frac{1}{2M^x}\nabla^2+\hat{U}_{\text{eff}}^x|\phi^x_j}, \\
    \mat S^x_{ij} := & \braket{\phi^x_i|\phi^x_j}, \\
    \mat B^x_{ij} := & \braket{\phi^x_i|\frac{\text{d}}{\text{d} t}\phi^x_j}, \\
    \matv B^x_{I,ij} := & \braket{\phi^x_i|\frac{\partial}{\partial \vect R^x_I}\phi^x_j},
\end{aligned}
\end{equation}
where the superscript $x=\{e,p\}$ represents either the electronic or protonic part of the multicomponent quantum system. 

\par Using these matrix notations, the action of the system is rewritten as 
\begin{equation}
\begin{aligned}
\label{eq:A_acb}
    A
    & = \int_{t0}^{t1}\text{d}t\int \text{d}\mathbf{k} \sum_n\sum_{ij} {c_{n\mathbf{k},i}^{e *}}\left[ (i\mat B^e_{ij} - \mat T^e_{ij})c_{n\mathbf{k},j}^e + i\mat S^e_{ij}\dot{c}_{n\mathbf{k},j}^e\right] - A^e_\text{pot} \\
    & \quad + \int_{t0}^{t1}\text{d}t \sum_n\sum_{ij} {c_{n,i}^{p *}}\left[ (i\mat B^p_{ij} - \mat T^p_{ij})c_{n,j}^p + i\mat S^p_{ij}\dot{c}_{n,j}^p\right] - A^p_\text{pot} - A^{ep}_\text{pot} \\
    & \quad + \int_{t0}^{t1}\text{d}t\sum_{I}^{N^c} \frac{1}{2}M_I \dot{\vect{R}}_{I}(t) ^2- A^c_\text{pot}, 
\end{aligned}
\end{equation}
with
\begin{equation*}
\begin{aligned}
    A^e_\text{pot} & =  \int_{t0}^{t1}\text{d}t \int \text{d}\vect{r}^e \left[ 
    - \frac{1}{2}V^e_\text{es}(\vect r^e, \vect R) 
    - V^c_\text{es}(\vect r^e, \vect R) \right]\rho^e(\vect{r}^e,t) + \int_{t0}^{t1}\text{d}tE_{XC}^e[\rho^e], \\
    A^p_\text{pot} & =  \int_{t0}^{t1}\text{d}t \int \text{d}\vect{r}^p \left[
    \frac{1}{2}V^p_\text{es}(\vect r^p, \vect R) 
    + V^c_\text{es}(\vect r^p, \vect R) \right]\rho^p(\vect{r}^p,t) + \int_{t0}^{t1}\text{d}tE_{XC}^p[\rho^p], \\
    A^{ep}_\text{pot} & = \int_{t0}^{t1}\text{d}t \int \text{d}\vect{r}^e 
    \left[- V^p_\text{es}(\vect r^e, \vect R)\right] 
    \rho^e(\vect{r}^e,t) + \int_{t0}^{t1}\text{d}t E_{epc}[\rho^e, \rho^p], \\
    A^c_\text{pot} & = \int_{t0}^{t1}\text{d}t \sum_{I<J}^{N^c}\frac{Z_I Z_J }{|\vect{R}_{I}(t)-\vect{R}_{J}(t) | },
\end{aligned}
\end{equation*}
where $\vect R$ represents all the classical degrees of freedom $\{\vect R_I\}$.
For the numerical implementation, the EOM 
for the electrons and quantum protons (Eqs. \ref{eq:tdks_e} and \ref{eq:tdks_p})
must be given in terms of the expansion coefficients $\{c_{n\mathbf{k},i}^{e*},c_{n,i}^{p*}\}$. This yields the TD-KS equations in the basis representation
\begin{equation}
\label{eq:EOM_wf}
\begin{aligned}
    \dot{c}_{n\mathbf{k},i}^e &= - \sum_{jl} {\mat S_{ij}^e}^{-1}(\mat B_{jl}^e + i\mat H_{jl}^e) c_{n\mathbf{k},l}^e = - i\sum_{I,jl} {\mat S_{ij}^e}^{-1}(\mat H_{jl}^e - i\dot{\vect R}^e_I\matv B_{I,jl}^e) c_{n\mathbf{k},l}^e, \\
    \dot{c}_{n,i}^p &= - \sum_{jl} {\mat S_{ij}^p}^{-1}(\mat B_{jl}^p + i\mat H_{jl}^p) c_{n,l}^p = - i\sum_{I,jl} {\mat S_{ij}^p}^{-1}(\mat H_{jl}^p - i\dot{\vect R}^p_I\matv B_{I,jl}^p) c_{n,l}^p.
\end{aligned}
\end{equation}
For the Euler-Lagrange equation (Eq. \ref{eq:EL_equation}) with respect to the classical nuclei,
\begin{equation}
\begin{aligned}
\label{eq:ELE1}
    \frac{\partial L}{\partial \dot{\vect R}_I(t)} & = M_I \dot{\vect{R}}_I(t) + i\sum_{n,ij} \left[\int \text{d}\mathbf{k}  {c_{n\mathbf{k},i}^e}^{*}\matv B^e_{I,ij}c_{n\mathbf{k},j}^e + {c_{n,i}^p}^{*}\matv B^p_{I,ij}c_{n,j}^p\right],\\
    \frac{\partial L}{\partial \vect R_I(t)} & = - \frac{\partial E_\text{tot}}{\partial \vect R_I(t)} 
    + i\sum_{n,ij} \frac{\partial}{\partial \vect R_I(t)}\left[ \int \text{d}\mathbf{k}{c_{n\mathbf{k},i}^e}^{*}\left(\mat B^e_{ij}c_{n\mathbf{k},j}^e + \mat S^e_{ij}\dot{c}_{n\mathbf{k},j}^e\right) 
    + {c_{n,i}^p}^{*}\left( \mat B^p_{ij}c_{n,j}^p + \mat S^p_{ij}\dot{c}_{n,j}^p\right)\right].
\end{aligned}
\end{equation}
Here, $E_\text{tot}$ is defined as the total energy of this quantum-classical NEO multicomponent system as
\begin{equation}
\begin{aligned}
    \label{eq:E_tot_1}
    E_\text{tot}
    & = \int \text{d}\vect{r}^{e} \int \text{d}\mathbf{k} \sum_n {\psi_{n\mathbf{k}}^e}^{*} (\vect{r}^e,t) \left[
    - \frac{1}{2}\nabla_{\vect{r}^e}^{2} 
    - V_{\text{es}}^{c}(\vect r^e) 
    - V_{\text{es}}^{p}(\vect r^e) - \frac{1}{2}V_{\text{es}}^{e}(\vect r^e)
    \right]\psi_{n\mathbf{k}}^e(\vect{r}^e,t) + E_{XC}^e[\rho^e] \\
    & + \int \text{d}\vect{r}^p \sum_n {\psi_n^{p}}^* (\vect{r}^p,t) \left[
    - \frac{1}{2M^p}\nabla_{\vect{r}^p}^{2}
    + V_{\text{es}}^{c}(\vect r^p) 
    + \frac{1}{2}V_{\text{es}}^{p}(\vect r^p)
    \right]\psi_n^{p}(\vect{r}^p,t) + E_{XC}^p[\rho^p] + E_{epc}[\rho^e, \rho^p] \\
    & 
    + \sum_{I}^{N^c} \frac{1}{2}M_I \dot{\vect{R}}_{I}^2(t)
    + \sum_{I<J}^{N^c}\frac{Z_I Z_J}{|\vect{R}_{I}(t)-\vect{R}_{J}(t) | }.
\end{aligned}
\end{equation}

\par The working equation for the numerical implementation of the Euler-Lagrange equation (Eq. \ref{eq:EL_equation})  
requires an explicit expression in the basis set representation for Eq. \ref{eq:ELE1}.
Using the proton-related terms as an example, the general expression is
\begin{equation}
\begin{aligned}
    \frac{\text{d}}{\text{d}t}\left[{c_{n,i}^p}^{*}\matv B^p_{I,ij}c_{n,j}^p \right]
    = & \dot{c}_{n,i}^p{}^{*}\matv B^p_{I,ij}c_{n,j}^p 
    + {c_{n,i}^p}^{*}\dot{\matv B}^p_{I,ij}c_{n,j}^p 
    + {c_{n,i}^p}^{*}\matv B^p_{I,ij}\dot{c}_{n,j}^p \\
    = & \sum_{ab} {c_{n,a}^p}^{*}{(i\mat H_{ab}^p - \mat B_{ab}^p )}{\mat S_{bi}^p}^{-1}\matv B^p_{I,ij}c_{n,j}^p 
    + {c_{n,i}^p}^{*}\dot{\matv B}^p_{I,ij}c_{n,j}^p \\
    & - {c_{n,i}^p}^{*}\matv B^p_{I,ij}\sum_{ab}{\mat S_{ja}^p}^{-1}(\mat B_{ab}^p + i\mat H_{ab}^p) c_{n,b}^p, \\
    \frac{\partial}{\partial \vect R_I}\left[{c_{n,i}^p}^{*}\left( \mat B^p_{ij}c_{n,j}^p + \mat S^p_{ij}\dot{c}_{n,j}^p\right) \right] 
    = & {c_{n,i}^p}^{*}\frac{\partial \mat B^p_{ij}}{\partial \vect R_I}c_{n,j}^p 
    + {c_{n,i}^p}^{*}\frac{\partial \mat S^p_{ij}}{\partial \vect R_I}\dot{c}_{n,j}^p \\
    = & {c_{n,i}^p}^{*}\frac{\partial \mat B^p_{ij}}{\partial \vect R_I}c_{n,j}^p
    - {c_{n,i}^p}^{*}({\matv B_{I,ij}^p}^* + \matv B_{I,ij}^p)\sum_{ab}{\mat S_{ja}^p}^{-1}(\mat B_{ab}^p + i\mat H_{ab}^p) c_{n,b}^p.\\
\end{aligned}
\end{equation}
Substituting Eq. \ref{eq:ELE1} into Eq. \ref{eq:EL_equation} and using expressions such as those given in the above equations, 
the EOM for the classical nuclei with atom-centered basis functions is

\begin{equation}
\begin{aligned}
\label{eq:EOM_c}
    M_I&\ddot{\vect R}_I (t)  =  - \frac{\partial E_\text{tot}}{\partial \vect R_I(t)} \\
    & + \sum_{n,ijlm}\int \text{d}\mathbf{k}  {c_{n\mathbf{k},i}^e}^{*}\left(
    { \matv B^e_{I,il}}^{*} {\mat S^e_{lm}}^{-1} \mat H^e_{mj} 
    + \mat H^e_{il}{\mat S^e_{lm}}^{-1} \matv B^e_{I,mj} 
    \right)c_{n\mathbf{k},j}^e \\
    & + \sum_{n,ijlm} {c_{n,i}^p}^{*}\left(
    { \matv B^p_{I,il}}^{*} {\mat S^p_{lm}}^{-1} \mat H^p_{mj} 
    + \mat H^p_{il}{\mat S^p_{lm}}^{-1} \matv B^p_{I,mj} 
    \right)c_{n,j}^p \\
    & + i\sum_{n,ijlm}\int \text{d}\mathbf{k}{c_{n\mathbf{k},i}^e}^{*}\left(
    { \mat B^e_{il}}^{*} {\mat S^e_{lm}}^{-1} \matv B^e_{I,mj} 
    - {\matv B^e_{I,il}}^{*} {\mat S^e_{lm}}^{-1} \mat B^e_{mj} 
    + \braket{\frac{\partial \phi^e_i}{\partial \vect R_I }|\frac{\text{d} \phi^e_j}{\text{d} t}} 
    - \braket{\frac{\text{d} \phi^e_i}{\text{d} t}|\frac{\partial \phi^e_j}{\partial \vect R_I}}
    \right)c_{n\mathbf{k},j}^e \\
    & + i\sum_{n,ijlm} {c_{n,i}^p}^{*}\left(
    { \mat B^p_{il}}^{*} {\mat S^p_{lm}}^{-1} \matv B^p_{I,mj} 
    - {\matv B^p_{I,il}}^{*} {\mat S^p_{lm}}^{-1} \mat B^p_{mj} 
    + \braket{\frac{\partial \phi^p_i}{\partial \vect R_I }|\frac{\text{d} \phi^p_j}{\text{d} t}} 
    - \braket{\frac{\text{d} \phi^p_i}{\text{d} t}|\frac{\partial \phi^p_j}{\partial \vect R_I}}
    \right)c_{n,j}^p. \\ 
\end{aligned}
\end{equation}
Eq. \ref{eq:EOM_c} includes the bra-kets with the dependence on the velocity of the basis functions, and they guarantee momentum conservation.\cite{kunert2003non}
At the same time, the last two summation terms with $i\sum_{n,ijlm}$ are usually omitted in practical implementations,\cite{doltsinis2002first,ojanpera2012nonadiabatic,hekele2021all} and we also follow this simplification in this work, retaining only the first three terms in Eq. \ref{eq:EOM_c} in our implementation. 
We refer to this approach as ``RT-NEO-TDDFT Ehrenfest dynamics''.

\subsection{RT-NEO-TDDFT  Ehrenfest Dynamics with Traveling Proton Basis} 
\label{sec:TPB}

\subsubsection{Traveling Proton Basis Approach} 
\label{subsec:TPB1}

\par In RT-NEO-TDDFT Ehrenfest dynamics, the classical nuclei move according to the EOM exerted by the time-dependent density in the context of RT-NEO-TDDFT.
Conventionally, the basis function centers for the electrons are centered on the atomic nuclei and therefore move with the nuclei. 
In principle, the basis function centers corresponding to quantum protons can stay stationary even in RT-NEO-TDDFT Ehrenfest dynamics.
When the quantum protons move far away from their original location, however, a single localized basis function center for each quantum proton is not sufficient. 
One strategy is to introduce multiple localized basis functions centered on fictitious ``ghost" atoms.\cite{zhao2020real,xu2022nuclear} 
However, this scheme is not always convenient because an approximate trajectory of the quantum proton needs to be known ahead of time to keep the number of extra proton basis function centers manageable. 

\par Alternatively, \citeauthor{zhao2020nuclear} proposed an approach called the traveling proton basis (TPB) such that fictitious classical degrees of freedom are added to propagate the basis function centers for the quantum protons. \cite{zhao2020nuclear}
In this original TPB approach, the equations of motion for the basis functions associated with the quantum protons are given by the energy gradient with respect to their basis function center positions using a fictitious proton mass.
This TPB idea entails extending the original NEO Lagrangian (Eq. \ref{eq:lagrangian}) as
\begin{equation}
L^\text{NEO-TPB}(t,\vect R,\dot{\vect R})=L^\text{NEO}(t,\vect R,\dot{\vect R})+ \sum_N\frac{1}{2}M^p {\dot{\vect R_N^p}^{2}},
\end{equation}
where $M^p$ is the fictitious mass of the proton basis function center and $\dot{\vect R_N^p}$ is the velocity associated with the basis function center $\vect R_N^p$.
For brevity, we omit the explicit dependence of the Lagrangian on the proton and electron KS wavefunctions here. 
Given the Lagrangian of the system, the classical EOM (analogous to Eq. \ref{eq:EOM_c}) is formally given by the Euler-Lagrange equation (Eq. \ref{eq:EL_equation}) which becomes
\begin{equation}
\label{eq:EOM_c_2}
\begin{aligned}
    M^p&\ddot{\vect R}_N^p (t)  =  - \frac{\partial E_\text{tot}}{\partial \vect R_N^p(t)} \\
    & + \sum_{n,ijlm}\int \text{d}\mathbf{k}  {c_{n\mathbf{k},i}^e}^{*}\left(
    { \matv B^e_{N,il}}^{*} {\mat S^e_{lm}}^{-1} \mat H^e_{mj} 
    + \mat H^e_{il}{\mat S^e_{lm}}^{-1} \matv B^e_{N,mj} 
    \right)c_{n\mathbf{k},j}^e \\
    & + \sum_{n,ijlm} {c_{n,i}^p}^{*}\left(
    { \matv B^p_{N,il}}^{*} {\mat S^p_{lm}}^{-1} \mat H^p_{mj} 
    + \mat H^p_{il}{\mat S^p_{lm}}^{-1} \matv B^p_{N,mj} 
    \right)c_{n,j}^p \\
    & + i\sum_{n,ijlm}\int \text{d}\mathbf{k}{c_{n\mathbf{k},i}^e}^{*}\left(
    { \mat B^e_{il}}^{*} {\mat S^e_{lm}}^{-1} \matv B^e_{N,mj} 
    - {\matv B^e_{N,il}}^{*} {\mat S^e_{lm}}^{-1} \mat B^e_{mj} 
   + \braket{\frac{\partial \phi^e_i}{\partial \vect R_N^p }|\frac{\text{d} \phi^e_j}{\text{d} t}} 
    - \braket{\frac{\text{d} \phi^e_i}{\text{d} t}|\frac{\partial \phi^e_j}{\partial \vect R_N^p}}
  \right)c_{n\mathbf{k},j}^e \\
   & + i\sum_{n,ijlm} {c_{n,i}^p}^{*}\left(
    { \mat B^p_{il}}^{*} {\mat S^p_{lm}}^{-1} \matv B^p_{N,mj} 
   - {\matv B^p_{N,il}}^{*} {\mat S^p_{lm}}^{-1} \mat B^p_{mj} 
   + \braket{\frac{\partial \phi^p_i}{\partial \vect R_N^p }|\frac{\text{d} \phi^p_j}{\text{d} t}} 
    - \braket{\frac{\text{d} \phi^p_i}{\text{d} t}|\frac{\partial \phi^p_j}{\partial \vect R_N^p}}\right)c_{n,j}^p, \\ 
\end{aligned}
\end{equation}
for the traveling proton basis function center $\vect R_N^p$.
In the original TPB scheme,\cite{zhao2020nuclear} the EOM was given by the derivative of the total energy and thus the four additional terms do not appear. 
As discussed above for Eq. \ref{eq:EOM_c}, the last two summation terms with $i\sum_{n,ijlm}$ are omitted in our implementation. 
In the following discussions, we refer to this Lagrangian-derived TPB scheme as l-TPB scheme. 
The pros and cons of the TPB and the new Lagrangian-derived l-TPB schemes are discussed using a numerical example in Sec. \ref{sec:TPB_compare}. 
In both cases, the fictitious mass of the proton basis function center is typically chosen such that the position expectation value corresponding to the quantum proton density and the proton basis function center closely follow each other. 
The physical proton mass is shown to be the correct mass for describing the dynamics of the expectation value of the proton position in a harmonic potential.\cite{zhao2021excited} 

\subsubsection{Alternative Formulation of TPB Approach} 
\label{subsec:TPB2}

\par 
In addition to the original TPB approach discussed above,\cite{zhao2020nuclear} herein we propose an alternative TPB approach by applying a semiclassical approximation to the Lagrangian instead of introducing an extra classical kinetic energy term for the quantum proton basis function center.
For a given proton KS orbital $n$, its translational degree of freedom can be expressed using the explicit phase with the momentum $\vect k_n^p$,
\begin{equation}
\label{eq:transf}
    \psi_n^{p} (\vect{r}^p,t)=  e^{i\vect k_n^p \cdot \vect r^p} \psi_n^{\prime p}(\vect{r}^p,t). 
\end{equation}
This does not alter the proton density (i.e., $\rho^p(\vect{r}^p,t) = {\rho^p}'(\vect{r}^p,t)$) nor the quantum dynamics in RT-TDDFT. 
With this rearranged expression, we have 
\begin{equation}
\begin{aligned}
    \bra{\psi_n^{p}}\frac{1}{2M^p}\nabla_{\vect r^p}^2 \ket{\psi_n^{p}} 
    =& \quad \bra{\psi_n^{\prime p}}\frac{1}{2M^p}\nabla_{\vect r^p}^2 \ket{\psi_n^{\prime p}} 
    - \frac{{\vect k_n^p}^2}{2M^p} 
    + i\frac{\vect k_n^p}{M^p} \bra{\psi_n^{\prime p}}\nabla_{\vect r^p}\ket{\psi_n^{\prime p}}
    \\
    \bra{\psi_n^{p}}i\frac{\text{d}}{\text{d} t} \ket{\psi_n^{p}} 
    =& \quad - \dot{\vect k_n^p}\bra{\psi_n^{\prime p}}\hat{\vect r}^{p}\ket{\psi_n^{\prime p}}  
    + \bra{\psi_n^{\prime p}}i\frac{\text{d}}{\text{d} t} \ket{\psi_n^{\prime p}} 
\end{aligned}
\end{equation}
Then the NEO Lagrangian (Eq. \ref{eq:lagrangian}) reads
\begin{equation}
\begin{aligned}
\label{eq:L_TPB2}
     L^\text{NEO} = &\quad {L'}^\text{NEO} 
    - \sum_{n,p}\left(\dot{\vect k_n^p}\bra{\psi_n^{\prime p}}\hat{\vect r}^{p}\ket{\psi_n^{\prime p}} 
    +\frac{{\vect k_n^p}^2}{2M^p}
    - i\frac{\vect k_n^p}{M^p}\bra{\psi_n^{\prime p}}\nabla_{\vect r^p}\ket{\psi_n^{\prime p}}\right) 
\end{aligned}
\end{equation}
where ${L'}^\text{NEO}$ has the same form as the NEO Lagrangian in Eq. \ref{eq:lagrangian} but expressed in terms of $\psi_n^{\prime p}$ instead of $\psi_n^{p}$ (see Eq. \ref{eq:transf}).

\par Next we make a semiclassical approximation to the translational degree of freedom associated with the proton KS orbital so that it can be treated separately as a classical degree of freedom in the RT-NEO-TDDFT Ehrenfest dynamics in the same spirit as in the original TPB approach.\cite{zhao2020nuclear}
This entails substituting $\vect k_n^p$ with the classical momentum $\vect k_n^p = M^p\dot{\vect R_N^p}$, 
where we assume that the proton orbital $\psi_n^{p}$ is localized on the proton basis function center $\vect R_N^p$.
In making this semiclassical approximation, we must take into account that the Lagrangian is defined with a negative sign for the quantum mechanical kinetic energy and with a positive sign for the classical mechanical kinetic energy.
Therefore, the sign of this kinetic energy term is changed as we make this semiclassical approximation to the translational kinetic energy associated with the proton KS orbitals. 
Additionally, this semiclassical approximation results in a term with $\ddot{\vect R_N^p}$ in the corresponding Lagrangian.
Because such a higher derivative term in the Lagrangian is known to lead to instability (i.e., Ostrogradsky instability \cite{motohashi2015third}), we simply neglect the $\ddot{\vect R_N^p}$ dependent term, $- M^p\ddot{\vect R_N^p}\bra{\psi_n^{p\prime}}\hat{\vect r}^p\ket{\psi_n^{p\prime}}$. 
The NEO Lagrangian under this semiclassical approximation is
\begin{equation}
\label{eq:L_neo-sc-tpb}
    L^\text{NEO-sc-TPB} = \quad {L'}^\text{NEO} 
    + \sum_N\left(\frac{1}{2}M^p {\dot{\vect R_N^p}^{2}} 
    - i\dot{\vect R_N^p}\sum_n\bra{\psi_n^{p\prime}}\nabla_{\vect R_N^p}\ket{\psi_n^{p\prime}}\right).
\end{equation}
In this new NEO-sc-TPB Lagrangian obtained by converting the translational degrees of freedom for the proton KS orbitals to classical degrees of freedom, the EOM for the proton KS orbital coefficients and the classical dynamics of the proton basis function centers can be derived via the Euler-Lagrange equation in the usual manner.
The EOM for the quantum proton degrees of freedom (corresponding to Eq. \ref{eq:EOM_wf}) is 
\begin{equation}
\label{eq:EOM_wf_3}
    \dot{c}_{n,i}^p = 
    -i \sum_{jl} {\mat S_{ij}^p}^{-1}\mat H_{jl}^p c_{n,l}^p.
\end{equation}
The EOM for the proton basis function centers, which are now classical degrees of freedom, is 
\begin{equation}
\begin{aligned}
\label{eq:EOM_c_3}
    M^p&\ddot{\vect R_N^p} (t)  =  - \frac{\partial E_\text{tot}}{\partial \vect R_N^p(t)} \\
    & + \sum_{n,ijlm}\int \text{d}\mathbf{k}  {c_{n\mathbf{k},i}^e}^{*}\left(
    { \matv B^e_{N,il}}^{*} {\mat S^e_{lm}}^{-1} \mat H^e_{mj} 
    + \mat H^e_{il}{\mat S^e_{lm}}^{-1} \matv B^e_{N,mj} 
   \right)c_{n\mathbf{k},j}^e \\
   & + i\sum_{n,ijlm}\int \text{d}\mathbf{k}{c_{n\mathbf{k},i}^e}^{*}\left(
   { \mat B^e_{il}}^{*} {\mat S^e_{lm}}^{-1} \matv B^e_{N,mj} 
   - {\matv B^e_{N,il}}^{*} {\mat S^e_{lm}}^{-1} \mat B^e_{mj} 
    + \braket{\frac{\partial \phi^e_i}{\partial \vect R_N^p }|\frac{\text{d} \phi^e_j}{\text{d} t}} 
    - \braket{\frac{\text{d} \phi^e_i}{\text{d} t}|\frac{\partial \phi^e_j}{\partial \vect R_N^p}}
   \right)c_{n\mathbf{k},j}^e. \\
\end{aligned}
\end{equation} 
Note that the terms with proton nonadiabatic couplings 
(the $B_{jl}^p$ term in Eq. \ref{eq:A_acb} and the $i\dot{\vect R_N^p}\sum_n\bra{\psi_n^{p\prime}}\nabla_{\vect R_N^p}\ket{\psi_n^{p\prime}}$ term in Eq. \ref{eq:L_neo-sc-tpb}) 
cancel each other for propagating both the proton wavefunctions and the proton basis function centers in this alternative formulation of the TPB approach. 
\par 
Let us now discuss the issue of energy conservation. 
For a given Lagrangian, the energy conservation is formally given by 
$\frac{dH}{dt}+\frac{\partial L}{\partial t}=0$, and the Hamiltonian is defined as
\begin{equation}
H\equiv \sum_i \dot{q}_i \frac{\partial L}{\partial \dot{q}_i} - L,
\end{equation}
where $q_i$ are the generalized coordinate variables. 
In mixed quantum-classical systems, we have $q_i=(\{\mathbf{R}_I\},\{\psi_i^*\})$. 
In most cases, the Lagrangian has no explicit dependence on time (i.e. $\partial_t L=0$) and the constant of motion is therefore given by the corresponding Hamiltonian.
However, in the RT-NEO-TDDFT approach, the term $i\partial_t (= \hat{H}^{KS}(t))$ in the Lagrangian (Eq. \ref{eq:lagrangian}) leads to $\partial_t L\neq0$ because of its explicit time dependence, i.e., for a general electronic or protonic orbital, $\bra{\psi_i(t)}\hat{H}^{KS}(t)\ket{\psi_i(t)}=\epsilon^{KS}_i(t)$. 
Thus, the conserved quantity is no longer given by the Hamiltonian alone.
For the NEO Lagrangian defined in Eq. \ref{eq:lagrangian}, the conserved quantity is given by
\begin{equation}
    \label{eq:cons}
    \frac{dH}{dt}+\frac{\partial L^{NEO}}{\partial t} = \frac{d}{dt}(E_\text{tot}+\sum_{N}\frac{1}{2}M^p {\dot{\vect R_N^p}^{2}}) = 0.
\end{equation}
In practical simulations, a few numerical simplifications are used, as discussed above, such as neglecting the velocity-dependent terms in Eq. \ref{eq:EOM_c}. 
The energy conservation is presented and discussed in Sec. \ref{sec:TPB_compare}.
The semiclassical approximation of the TPB approach discussed in this section is referred to as the sc-TPB approach for simplicity.

\subsection{Numerical Implementation}
\subsubsection{Energy Derivatives}
\label{subsec:F_NEODFT}
\par
In our previous work,\cite{xu2022nuclear} the total energy for multicomponent NEO-DFT (as defined by Eq. \ref{eq:E_tot_1}) is calculated with a NEO-induced correction to the conventional DFT energy,
\begin{equation}
\begin{aligned}
\label{eq:NEO_Etot}
     E^\text{NEO}_\text{tot}=&\sum_{l}f_l\epsilon_l^\text{NEO}-\int v_{xc}^{e-\text{NEO}}(\vect r^e)\rho^e(\vect r^e)\text d\vect r^e+E_{xc}^{e-\text{NEO}}[\rho^e(\vect r^e)]-\\ &\frac{1}{2}\int v_\text{es}^{e-\text{NEO}}{(\vect r^e)}\rho^e(\vect r^e)\text d\vect r^e +E_\text{nuc-nuc}+\Delta E^\text{NEO}_\text{nuc-nuc},
\end{aligned}
\end{equation} 
where $\epsilon_l^{\text{NEO}}$ are the KS eigenvalues from the NEO electron KS Hamiltonian. 
$v_{xc}^{e-\text{NEO}}$ and $E_{xc}^{e-\text{NEO}}$ are the exchange-correlation  potential and energy of the NEO electron KS Hamiltonian,
which includes both electron-electron exchange-correlation and electron-proton correlation.
The fourth term with $v_\text{es}^{e-\text{NEO}}$ corrects for the double counting in the electrostatic interactions.
$E_\text{nuc-nuc}$ is the energy contributed from interactions between classical nuclei, as in conventional DFT.
The energy terms involving the quantum protons take the form
\begin{equation}
\label{eq:dE_NEO}
    \Delta E^\text{NEO}_\text{nuc-nuc}=T^p+ \frac{1}{2}\left(J^p-K^p\right) - E^{cpp} + \sum_I^\text{nuclei} Z_I\left[V_\text{es}^{p}(\vect R_I)-V_\text{es}^{cp}(\vect R_I)\right],
\end{equation}
where $T^p,J^p$, and $K^p$ are the kinetic, Coulomb, and exchange energies of the quantum protons, and
$E^{cpp}$ is the electrostatic energy between (quantum) protons in conventional DFT calculations to remove double counting.
In the last term, $V_\text{es}^{p}$ represents the Coulomb interaction between the quantum protons and the classical nuclei $I$ with charge $Z_I$, and $V_\text{es}^{cp}$ is the analogous term in conventional DFT calculations to remove double counting.
RT-NEO-TDDFT Ehrenfest dynamics requires the evaluation of the partial derivative of the NEO total energy $E^\text{NEO}_\text{tot}$ with respect to the classical nuclear coordinate $\vect R_I$ (see Eq. \ref{eq:EOM_c}), defining the force on nucleus $I$ as
\begin{equation}
\vect F_I= -\frac{\partial }{\partial \vect R_I}E^\text{NEO}_\text{tot} = -\frac{\partial }{\partial \vect R_I}(E^\text{e-NEO} + \Delta E^\text{NEO}_\text{nuc-nuc}).
\end{equation}
This force can be classified as having an electron contribution and a quantum proton contribution. 

\par The electron contribution $\vect F^e_I = - {\partial E^\text{e-NEO}}/{\partial \vect R_I}$ has the same form as the conventional DFT force.
For example, in the FHI-aims code,\cite{blum2009ab} the forces are expressed as having several components including 
Hellmann-Feynman forces $\vect F^\text{HF}_I$, 
Pulay forces $\vect F^\text{P}_I$, 
electrostatic multipole derivatives $\vect F^\text{MP}_I$, 
scalar-relativistic derivatives $\vect F^\text{at.ZORA}_I$, and so forth.  
In our periodic NEO-DFT implementation, these terms are evaluated by the FHI-aims package in the same manner as for conventional DFT, but using the updated electron density $\rho^e$ and effective potential $v_\text{eff}^e$ that include the NEO effects.

\par The quantum proton contribution of  $\vect F^p_I = - {\partial \Delta E^\text{NEO}_\text{nuc-nuc}}/{\partial \vect R_I}$ needs to be explicitly evaluated for the individual terms in $\Delta E^\text{NEO}_\text{nuc-nuc}$ (Eq. \ref{eq:dE_NEO}).
A combination of numerical real-space grid integration and analytical integration is used to evaluate the proton Hamiltonian as discussed in Ref. \onlinecite{xu2022nuclear}, 
and the energy gradients are also computed accordingly.
For a given quantum proton basis function center $\vect R_N^p$, the gradient of the kinetic  energy $T^p$ and exchange energy $K^p$ (see Eq. \ref{eq:dE_NEO}) are evaluated with analytical integrals,
\begin{align}
\label{eq:F_T}
    \frac{\partial T^p}{\partial \vect R_N^p} = & \sum_{ij}\left[\frac{\partial \mat{D}_{ij}^p}{\partial \vect R_N^p} \mat{T}_{ij}^p + \mat{D}_{ij}^p\frac{\partial \mat{T}_{ij}^p}{\partial \vect R_N^p}\right], \\
    \frac{\partial K^p}{\partial \vect R_N^p} = & \sum_{ijkl}\left[2\frac{\partial \mat{D}_{ij}^p}{\partial \vect R_N^p}\left(ik|jl\right)\mat{D}_{kl}^p + 
    \mat{D}_{ij}^p\frac{\partial\left(ik|jl\right)}{\partial \vect R_N^p}\mat{D}_{kl}^p\right],
\end{align}
where
$\mat{D}_{ij}^p$ is the density matrix for the quantum protons, 
$\mat{T}_{ij}^p$ is the proton kinetic energy matrix defined as in Eq. \ref{eq:mat},
and $\left(ik|jl\right)$ is the `two electron integral'. 
All the above stated analytical integrals are evaluated using the libcint package. \cite{10.1002/jcc.23981}
The electrostatic energy is evaluated with numerical integrals.
For faster convergence with respect to distance,\cite{xu2022nuclear} the electrostatic potential of a quantum proton orbital $\psi_n^{p}$ and the corresponding classical proton at $\vect R_N^p$ are always evaluated together as a correction potential $V_{\text{es-corr}}^p\left(\vect R_N^p;\vect r \right) = V_{\text{es},N}^p\left(\vect R_N^p - \vect r \right) - {1}/{|\vect R_N^p - \vect r|}$, with $V_{\text{es},N}^p$ denoting the electrostatic potential of the quantum proton centered on $\vect R_N^p$.

\par As a part of $E^\text{NEO}_\text{nuc-nuc}$ (Eq. \ref{eq:dE_NEO}), we define $E^p_\text{es}$ as the sum of
the electrostatic energies contributed by the quantum protons and the classical protons 
\begin{equation}
\label{eq:E_qp-cp}
    E^p_\text{es} = \frac{1}{2}J^p-E^{cpp} = \frac{1}{2}\int d\vect r \rho^p(\vect r)\sum_N V_{\text{es-corr}}^p\left(\vect R_N^p;\vect r \right) + \frac{1}{2}\sum_{N,N'}V_{\text{es-corr}}^p\left(\vect R_N^p;\vect R_{N'}^p \right).
\end{equation}
The gradient of  the above energy contribution $E^p_\text{es}$ can be evaluated as 
\begin{equation}
\begin{aligned}
    \frac{\partial E^p_\text{es} }{\partial \vect R_N^p} = & 
    \sum_{ij,N'} \frac{\partial \mat{D}_{ij}^p}{\partial \vect R_N^p} \int d\vect r \phi^p_i(\vect r) \phi^p_j(\vect r) V_{\text{es},N'}^{p}\left( \vect R_{N'}^p - \vect r \right) + 
    \sum_{ij,N'} \mat{D}_{ij}^p \int d \vect r \phi^p_i(\vect r) \frac{\partial \phi^p_j(\vect r)}{\partial \vect R_N^p} V_{\text{es-corr}}^{p}\left(\vect R_{N'}^p;\vect r \right)
     \\
    & + \frac{1}{2}\int d \vect r \rho^p \left(\vect r \right) \left.\frac{\partial V_{\text{es-corr}}^p\left(\vect R_N^p;\vect r \right)}{\partial \vect R_N^p}\right|_{D^p} +
    \frac{1}{2} \sum_{N'\neq N} \frac{\partial V_{\text{es-corr}}^{p}\left(\vect R_{N'}^p;\vect R_N^p \right)}{\partial \vect R_N^p}.
\end{aligned}
\end{equation}
The gradient of the correction potential ${\partial V_{\text{es-corr}}^p}/{\partial \vect R_N^p}$ is evaluated numerically on real space grids using a multipole expansion.\cite{blum2009ab}
Similarly, the gradient of the electrostatic energy between the quantum protons and the classical nuclei, i.e., the last term in Eq. \ref{eq:dE_NEO}, can be written as: 
\begin{equation}
    \frac{\partial E_\text{nuc}^p}{\partial \vect R_N^p} =
    \sum_{ij,I} \frac{\partial \mat{D}_{ij}^p}{\partial \vect R_N^p} \int d\vect r \phi^p_i(\vect r) \phi^p_j(\vect r) \frac{Z_I}{\left| \vect R_N^p - \vect R_I \right|} +
    \sum_I Z_I \left.\frac{\partial V_{\text{es-corr}}^{p}\left(\vect R_N^p;\vect R_I \right)}{\partial \vect R_N^p}\right|_{\mat D^p}.
\end{equation}
For the classical nuclei $I$, the gradients of the first four terms in Eq.\ref{eq:dE_NEO} are zero, leading to
\begin{equation}
    \frac{\partial \Delta E^\text{NEO}_\text{nuc-nuc}}{\partial \vect R_I} = \frac{\partial E_\text{nuc}^p}{\partial \vect R_I} =
    Z_I\sum_N \frac{\partial V_{\text{es-corr}}^{p}\left(\vect R_N^p;\vect R_I \right)}{\partial \vect R_I}.
\end{equation}

As discussed in our recent work, \cite{xu2022nuclear} the electrostatic potential for the quantum proton $V_\text{es}^p$ and electron-proton correlation potential $V_\text{epc}$ are passed to the electron Hamiltonian $\hat{H}^e$ in the numerical implementation for evaluating the total energy.
Thus, $E_\text{es}^e$ and $E_\text{epc}$ are evaluated as a part of the electron contribution $E^\text{e-NEO}$ in Eq. \ref{eq:NEO_Etot}.
Because these two terms have contributions from both the electrons and the quantum protons, their energy gradients with respect to both the electron and proton degrees of freedom need to be addressed separately.
Although the electronic degree of freedoms are straightforward to evaluate within the FHI-aims code, 
the proton degrees of freedom need to be taken into account explicitly by calculating their partial gradients with respect to the proton basis function center coordinates as
\begin{align}
    \frac{\partial E_\text{es}^e}{\partial \vect R_N^p} = & 
    \sum_{ij} \frac{\partial \mat{D}_{ij}^p}{\partial \vect R_N^p} \int d\vect r \phi^p_i(\vect r) \phi^p_j(\vect r) V_\text{es}^{e}\left(\vect r \right) + 
    \int d \vect r \rho^e \left(\vect r \right) \left.\frac{\partial V_{\text{es-corr}}^p\left(\vect R_N^p;\vect r \right)}{\partial \vect R_N^p}\right|_{\mat D^p},\\
    \label{eq:F_epc}
    \frac{\partial E_\text{epc}}{\partial \vect R_N^p} = & 
    \sum_{ij} \frac{\partial \mat{D}_{ij}^p}{\partial \vect R_N^p} \int d\vect r \phi^p_i(\vect r) \phi^p_j(\vect r) V_\text{epc}^{p}\left(\vect r \right) + 
    \sum_{ij} 2\mat{D}_{ij}^p \int d\vect r \phi^p_i(\vect r) \frac{\partial \phi^p_j(\vect r)}{\partial \vect R_N^p} V_\text{epc}^{p}\left(\vect r \right),
\end{align}
where $V_\text{es}^{e}\left(\vect r \right)$ is the electrostatic potential from all electrons, and $V_\text{epc}^{p}\left(\vect r \right)$ 
is the electron-proton correlation potential for the quantum protons.

\par In summary, Eqs.~\ref{eq:F_T}-\ref{eq:F_epc} provide all the necessary expressions for evaluating the total energy gradient within the RT-NEO-TDDFT scheme.
Note that, unlike the total derivative, the partial derivative $\sum_{ij}{\partial \mat{D}_{ij}^p}/{\partial \vect R_N^p} = 0$, because the proton density matrix does not have an explicit dependence on the coordinates of the quantum proton basis function center, and this simplifies several equations for implementation.
See the Appendix for further information relevant to the ground-state NEO-DFT calculations.

\subsubsection{TD-KS Orbitals in RT-NEO-TDDFT Ehrenfest Dynamics}
\label{subsec:RT_propagation}

\par 
As discussed in Sec. \ref{sec:RT-NEO-TDDFT_L}, 
the dynamics of the electron and proton KS wave functions are governed by the coupled TD-KS equations within the RT-NEO-TDDFT framework,
\begin{align}
\label{eq:tdks_e2}
     i\frac{\text d}{\text dt} \psi^e_{i,\bf{k}} (\vect r^e,t)= & \hat{H}^e \psi^e_{i,\bf{k}} (\vect r^e,t) = \left[-\frac{1}{2}\nabla^2+U_{\text{eff}}^e(\vect r^e,t)+ \hat{v}_\text{ext}(t)\right] \psi^e_{i,\bf{k}}(\vect r^e,t) \\
 \label{eq:tdks_p2}
     i \frac{\text d}{\text dt}  \psi^p_i (\vect r^p,t) = & \hat{H}^p \psi^p_i (\vect r^p,t) = \left[-\frac{1}{2M^p}\nabla^2+U_{\text{eff}}^p(\vect r^p,t)-\hat{v}_\text{ext}(t)\right] \psi^p_i (\vect r^p,t).
\end{align}
where $U_{\text{eff}}$ is the effective potential that includes the Hartree potential, the exchange-correlation potential, and the electron-proton correlation potential as shown in Eq. \ref{equ:v_eff_e}. 
Although the exchange-correlation potentials depend on all former time in principle, we adopt the adiabatic approximation \cite{lacombe2023non} that ignores this history dependence and thus removes the explicit time dependence of $U_\text{eff}$.
Because the effective potential contains coupled contributions from the electrons and protons,
the electron and proton orbitals need to be propagated at the same time.
When needed, $\hat{v}_\text{ext}(t)$ can describe the (time-dependent) external electromagnetic field, which can be represented either in the length gauge $\hat{v}_\text{ext}(t)=\hat{\vect r} \cdot \vect E(t)$ or in the velocity gauge $\hat{v}_\text{ext}(t)=-i\vect A(t)\cdot \nabla + \vect A(t)^{2}$ with $\vect E(t)=-\partial_t \vect A(t)$ denoting the external electric field and $\vect A(t)$ denoting the corresponding vector potential. 

\par The electron KS wave functions are propagated according to the electronic TD-KS equation using the existing RT-TDDFT module in FHI-aims \cite{hekele2021all} with the NEO-modified effective potential as described in Ref. \onlinecite{xu2023first}.
The electronic effective potential is updated on-the-fly following the NEO-DFT approach\cite{xu2022nuclear} to include NEO contributions.
The proton degrees of freedom are propagated as follows in the RT-NEO-TDDFT Ehrenfest dynamics. 
As described in Eqs. \ref{eq:basis1} and \ref{eq:basis2} in Sec. \ref{sec:RT-NEO-TDDFT_B}, the time-dependent proton KS orbitals are expanded using a set of periodic Gaussian-type basis functions, each of which are associated with a center $I$.
Using the matrix notation, the equation of motion of the quantum protons in terms of the coefficients $c^p_{n,i}$ are
\begin{equation}
\label{eq:eom_p_rttddft}
    \frac{\text d}{\text dt} \bmat C(t) = -i\bmat{S}^{-1}\bmat H(t)\bmat C(t),
\end{equation}
where $\bmat S$ and $\bmat H$ are the overlap matrix and Hamiltonian matrix as defined in Eq. \ref{eq:mat}.
This differential equation can be solved using an exponential integration scheme, i.e., by propagating the proton KS wave functions with a given time step $\Delta t$: 
$\bmat C(t+\Delta t) = \text{exp}(\hat O) \bmat C(t).$\cite{magnus1954exponential}
The $\hat O$ here is any generic quantum mechanical operator.
With the Exponential-Midpoint (EM) propagator,\cite{gomez2018propagators} the time-dependent expansion coefficients are evolved as
\begin{equation}
\label{eq:eom_p_rttddft_EM}
    \bmat C(t+\Delta t) = \bmat S^{-\frac{1}{2}}\text{exp}\left(-i\Delta t \bmat S^{-\frac{1}{2}} \bmat H(t+\frac{\Delta t}{2}) \bmat S^{-\frac{1}{2}} \right) \bmat S^{\frac{1}{2}} \bmat C(t).
\end{equation}
Since the TD-KS equations are a set of non-linear equations, where the Hamiltonian depends on the KS wave functions, an iterative predictor/corrector method\cite{zhu2018self} is used to obtain the Hamiltonian matrix at time $t + \frac{\Delta t}{2}$.
The matrix exponential is computed by the eigenvectors method, \cite{moler1978nineteen} 
\begin{equation}
    \text{exp}\left(\bmat A\right) = \bmat V \text{diag}\left(e^{\lambda_1},...e^{\lambda_n}\right) \bmat V^{-1},
\end{equation}
where $\bmat V$ and $e^{\lambda_i}$ are the eigenvectors and eigenvalues, respectively, of matrix $\bmat A$.
Alternative approaches, such as the explicit fourth-order Runge–Kutta method, are also implemented for solving Eq. \ref{eq:eom_p_rttddft}.

\par In Ehrenfest dynamics, the nonadiabatic couplings between the quantum electrons/protons and the classical nuclei need to be computed.
With the TPB approaches, the positions of the classical nuclei, as well as the positions of the basis function centers for the quantum protons, are propagated on the fly as discussed in Sec. \ref{sec:TPB}.
The electron and proton KS wave functions are propagated according to the EOM in Eq. \ref{eq:EOM_wf} and/or \ref{eq:EOM_wf_3}, depending on the particular TPB scheme chosen.
In the case of the sc-TPB scheme, the EOM for the quantum proton is given by Eq. \ref{eq:EOM_wf_3}.
In the case of the TPB and l-TPB schemes, 
the nonadiabatic coupling (NAC) term is needed for the quantum proton propagation,\cite{zhao2020nuclear} as given by Eq. \ref{eq:EOM_wf}, yielding 
\begin{equation}
\label{eq:eom_p_rttddft_NAC}
    \frac{\text d}{\text dt} \bmat C(t) = -i\bmat{S}^{-1}[\bmat H(t)-i\sum_I\dot{\vect R_I}(t)\bmatv B_I]\bmat C(t),
\end{equation}
where $\bmatv B_I$ is the NAC matrix defined in Eq. \ref{eq:mat}.
Because of the anti-Hermitian nature of the $\bmatv B_I$ matrix, adding the NAC term does not change the hermiticity when evaluating the matrix exponential in Eq. \ref{eq:eom_p_rttddft_EM}.
The classical EOM for propagating the classical nuclei and the quantum proton basis function centers is given by Eq. \ref{eq:EOM_c_2} for the TPB and l-TPB schemes or Eq. \ref{eq:EOM_c_3} for the sc-TPB scheme. 
As discussed above, some of the terms are neglected in our implementation. 
In our Ehrenfest dynamics implementation, the positions of the classical nuclei and quantum proton basis function centers are propagated using the velocity Verlet method,\cite{verlet1967computer}
and all NEO contributions are taken into account as an additional effective Newtonian force on top of the conventional Ehrenfest forces as described in Sec. \ref{subsec:F_NEODFT}.

Algorithm \ref{alg:Ehrenfest} outlines the numerical implementation scheme for the NEO Ehrefest dynamics.
Because of the heavier mass of the quantum proton, the time step of the proton KS wave function propagation $\Delta t_\text{wf}^p$ can be larger than that associated with the electrons $\Delta t_\text{wf}^e$.
Similarly, the classical nuclei can be propagated with an even larger time step $\Delta t_\text{nuc}$.

\section{Results and Discussion}
\label{sec:result}
The above-described RT-NEO-TDDFT with Ehrenfest dynamics using the TPB schemes is implemented in the FHI-aims package.\cite{blum2009ab}
We first validated our implementation against the existing implementation in the Q-Chem code\cite{epifanovsky2021software} using isolated systems (see Supplementary Material).
For all the RT-NEO-TDDFT calculations discussed in this section, we used the PBE exchange-correlation  functional\cite{perdew1996generalized} for electrons, the exact exchange for quantum protons, and the epc17-2 functional\cite{yang2017development,brorsen2017multicomponent} for electron-proton correlation.
We used the intermediate numerical atomic orbital (NAO) basis set\cite{blum2009ab} for electrons
and the PB4-F2 basis set\cite{yu_development_2020} for the quantum protons.
All NEO-Ehrenfest dynamics simulations used the same electronic/protonic timestep, $\Delta t_\text{wf}^p = \Delta t_\text{wf}^e = 0.0048$ fs, and a larger timestep for the classical nuclei, $\Delta t_\text{nuc}=0.048$ fs.

\begin{algorithm}[H]
    \caption{NEO Ehrenfest dynamics implemented in this work}
    \label{alg:Ehrenfest}
\begin{algorithmic}
    \Require 
        $\Delta t_\text{wf}^e$, 
        $\Delta t_\text{wf}^p = N^p * \Delta t_\text{wf}^e$, 
        $\Delta t_\text{nuc} = N^c * \Delta t_\text{wf}^e$

\For {$i=1,2,...,$} \Comment{Ehrefnest step}
    \For {$j=1,2,...,N^c$} \Comment{Proton step}
        \State Propagate proton wavefucntion $\psi^p(t) \rightarrow \psi^p(t + \Delta t_\text{wf}^p)$ based on Eq. \ref{eq:EOM_wf} or \ref{eq:EOM_wf_3}
        \For {$k=1,2,...,N^p$} \Comment{Electron step}
            \State Propagate electron wavefucntion $\psi^e(t) \rightarrow \psi^e(t + \Delta t_\text{wf}^e)$ based on Eq. \ref{eq:EOM_wf}
            \State $t=t+\Delta t_\text{wf}^e$
        \EndFor
    \EndFor
     \State Propagate classical nuclei $\vect R_I(t) \rightarrow \vect R_I(t+\Delta t_\text{nuc})$ based on Eq. \ref{eq:EOM_c}
    \If{Traveling proton basis}
        \State Propagate proton basis function centers $\vect R_N^p(t) \rightarrow \vect R_N^p(t+\Delta t_\text{nuc})$ based on Eq. \ref{eq:EOM_c_2} or \ref{eq:EOM_c_3}
    \EndIf
    \State Update (numerical/analytical) atomic integrals based on new coordinates
\EndFor
\end{algorithmic}
\end{algorithm}

\subsection{Traveling Proton Basis schemes in RT-NEO-TDDFT Ehrenfest Dynamics}
\label{sec:TPB_compare}

\begin{figure}[ht]
    \centering
    \includegraphics[width=0.98\textwidth]
    {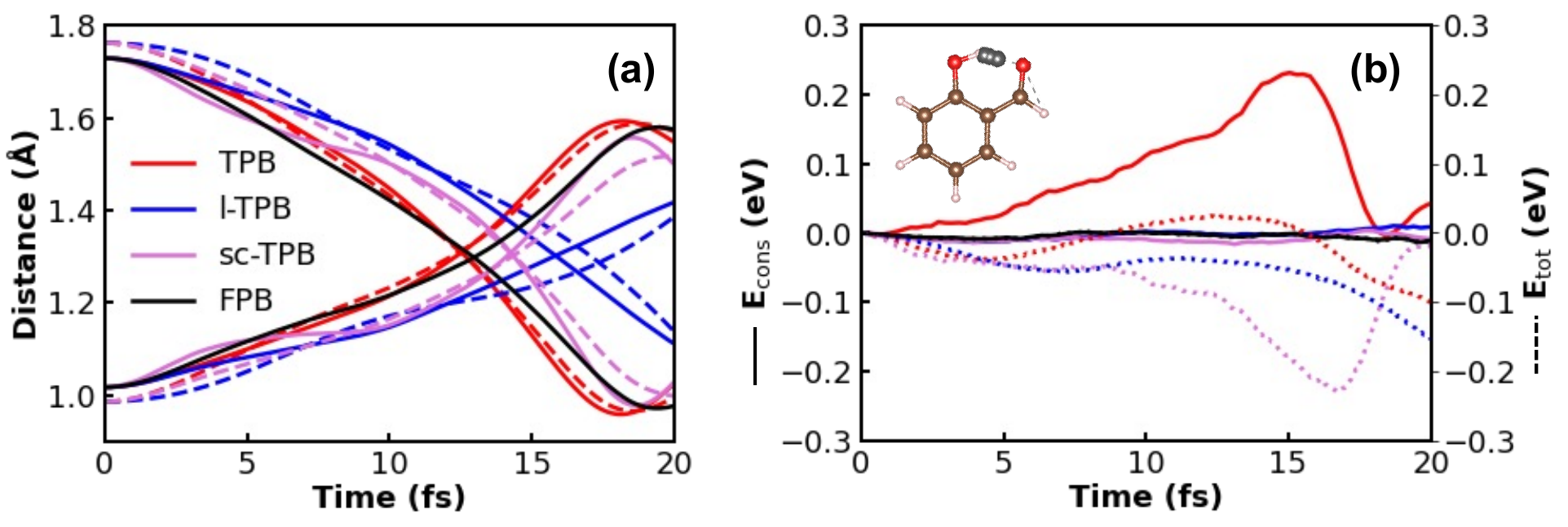}
    \caption{(a) The distances between the position expectation value of the transferring proton and the donor and acceptor oxygen atoms in the oHBA molecule as a function of time with the TPB (red), l-TPB (blue), sc-TPB (orchid), and FPB (black) approaches. The solid and dashed lines represent the expectation value of the quantum proton position operator and the proton basis function center, respectively. (b) Change of $E_\text{cons}$ (solid lines) and $E_\text{tot}$ (dotted lines) along the NEO Ehrenfest dynamics trajectory with the three TPB approaches and the FPB approach.
    $E_\text{cons}$, which was determined to be the conserved quantity for the 1-TPB and sc-TPB schemes based on the NEO Lagrangian, is the sum of the total system energy $E_\text{tot}$ and the kinetic energy of the proton basis function centers. The black solid and dotted lines are identical because the kinetic energy of the proton basis function centers is zero for the FPB approach. 
    The inset shows the molecular structure of the oHBA molecule with the grey sphere representing the position of additional 'ghost' proton basis function centers for the FPB approach.}
    \label{fig:mol_oHBA}
\end{figure}

\par We discuss here the performance of the three TPB schemes discussed in Sec. \ref{sec:TPB} by comparing to the standard RT-NEO-TDDFT Ehrenfest dynamics with fixed proton basis function centers (FPB) on ``ghost" atoms. 
In the limit of having enough basis function centers to represent any quantum proton dynamics, the FPB scheme is considered the most rigorous formulation of RT-NEO-TDDFT Ehrenfest dynamics. 
However, in practice, the FPB scheme can be used only with a finite number of proton basis function centers along a pre-defined spatial region of expected quantum proton dynamics. 
In the following, we use the results from the sc-TPB simulation to choose the locations of the proton basis function centers for the FPB simulation. 
In Supplementary Material, we present the FPB simulation results with the proton basis function centers placed according to the original TPB simulation,\cite{zhao2020nuclear} but no noticeable differences are found.

\par Following previous work,\cite{zhao2021excited} the excited state intramolecular proton transfer (ESIPT) process in an o-hydroxybenzaldehyde (oHBA) molecule was used for the comparison of different TPB schemes.
The molecule was initially optimized at the ground-state geometry.
Here, only the transferring proton is treated quantum mechanically using the NEO method.
At $t=0$, one electron was ``manually" promoted from the highest occupied molecular orbital (HOMO) to the lowest unoccupied molecular orbital (LUMO) to induce the ESIPT process.
Figure. \ref{fig:mol_oHBA} (a) shows the trajectories of the ESIPT of the oHBA molecule using the TPB approaches and the standard FPB approach. 
In Figure \ref{fig:mol_oHBA} (a), the solid lines and dashed lines are the expectation value of the quantum proton position operator and the location of the proton basis function center (for TPB schemes), respectively.
For simplicity, the proton transfer is considered to have occurred when the expectation value of the proton position operator is the same distance from the donor and acceptor oxygen atoms, i.e., when the two lines cross in Figure \ref{fig:mol_oHBA} (a).
Qualitatively, all TPB and FPB approaches show the proton transfer process on a similar time scale.
Quantitatively, the standard TPB scheme shows the best agreement with the reference FPB approach.
The sc-TPB scheme is somewhat worse, and the l-TPB scheme predicts a transfer time that is $\sim20\%$ longer.  
With the l-TPB scheme, the position of the proton basis function center (dashed blue lines) tends to drag behind the quantum proton position expectation value (the solid blue lines), and this artificial friction somewhat slows down the proton transfer.

\par Another aspect of examining the dynamics from the mathematical viewpoint is the conservation of the constant of motion for the dynamics governed by the Lagrangian.  
The conserved quantity for the l-TPB and sc-TPB approaches is given by Eq. \ref{eq:cons}. 
We will denote this quantity as $E_\text{cons}$, which is the sum of the total system energy $E_\text{tot}$ and the kinetic energy of the proton basis function centers. 
Note that we have $E_\text{tot}$ = $E_\text{cons}$ for the FPB approach since the kinetic energy of the proton basis function centers is zero. 
The conserved quantity for the original TPB approach\cite{zhao2020nuclear} was presumed to be $E_\text{tot}$ because the moving basis function center is not part of the physical system but rather is just a mathematical tool to allow the use of a smaller basis set. 
Figure \ref{fig:mol_oHBA} (b) shows $E_\text{cons}$ and $E_\text{tot}$ during the real-time propagation using the FPB and TPB approaches.
Having been derived directly from the NEO Lagrangian, both the l-TPB and sc-TPB schemes show stable behavior for $E_\text{cons}$, similar to the FPB scheme. 
None of the TPB approaches preserve the total energy $E_\text{tot}$ as well as the FPB approach, as discussed in previous work.\cite{zhao2020nuclear}

\par The TPB and sc-TPB schemes are appealing for different reasons. 
The original TPB scheme shows better performance in terms of how closely it agrees with the quantum dynamics of the reference FPB simulation, as determined by the motion of the expectation value of the proton position operator. 
At the same time, the sc-TPB scheme addresses the previously noted issue of energy conservation,\cite{zhao2020nuclear} and the expected conserved quantity is identified by deriving the dynamics from the NEO Lagrangian. 
In the most rigorous approach, as when a very accurate description of the quantum dynamics with energy conservation is necessary, the FPB scheme can be performed with pre-defined proton basis function center positions based on simulations with the original TPB or sc-TPB schemes. 

\subsection{Applications with Periodic Boundary Conditions}

\begin{figure}[ht]
    \centering
    \includegraphics[width=0.98\textwidth]{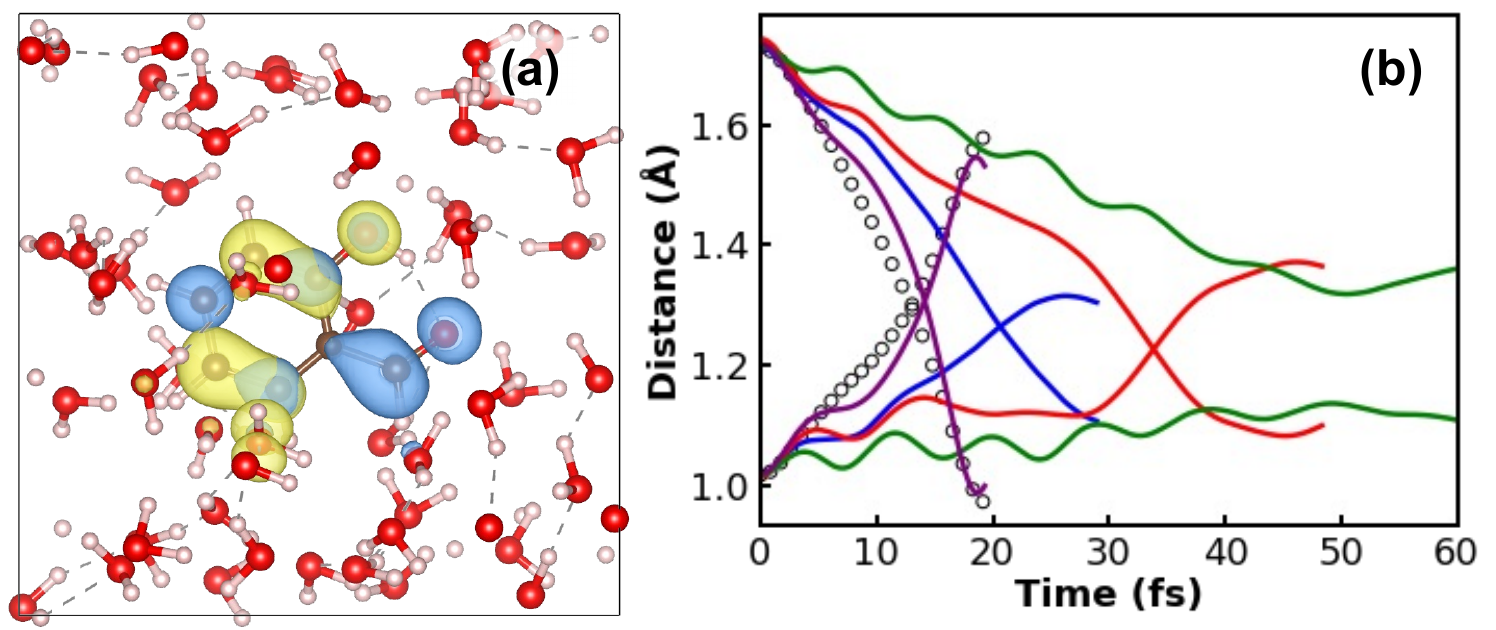}
    \caption{(a) A snapshot of the oHBA molecule solvated in liquid water from FPMD simulation. The isosurfaces show the molecular HOMO state (yellow) and the molecular LUMO state (blue). (b) The distances between the position expectation value of the transferring proton and the donor and acceptor oxygen atoms in the oHBA molecule as a function of time using the sc-TPB approach. The circles represent the isolated oHBA molecule in vacuum, and the colored solid lines represent the trajectories started from four randomly selected snapshots from the FPMD simulation.}
    \label{fig:water_traj}
\end{figure}

\par 
Finally, we demonstrate the utility of the newly developed extension of RT-NEO-TDDFT Ehrenfest dynamics with periodic boundary conditions (PBCs). 
In particular, we study how the explicit dynamics of solvating water molecules can significantly impact the ESIPT process. 
For this purpose, we study the ESIPT of the oHBA molecule that is solvated in explicit liquid water molecules.
To obtain equilibrium configurations of the oHBA molecule in water, we performed first-principles molecular dynamics (FPMD) simulations in a periodic cubic simulation cell containing the oHBA molecule and 57 water molecules at room temperature (300 K).\cite{xu2023first}
The SCAN meta generalized gradient approximation\cite{sun2015strongly} to the XC functional was used for FPMD, as it has been shown to provide a reasonably accurate structure of water.\cite{xu2019first}
The oHBA molecule was fixed at its ground state geometry during the FPMD simulations so that the same starting oHBA geometry could be used for all subsequent Ehrenfest dynamics simulations to enable the study of the effects from the dynamics of the solvating water molecules.

\par We then performed NEO Ehrenfest dynamics simulations using the  TPB and sc-TPB approaches on four randomly selected snapshots from the FPMD trajectory (Fig. \ref{fig:water_traj}(a)). 
The transferring proton in the oHBA molecule and all nearest protons of the surrounding H$_2$O molecules that exhibit hydrogen bonding with the two oxygen atoms of the oHBA molecule in the initial configuration were treated as quantum protons using NEO.
At $t=0$, a single electron was excited from the molecular HOMO to the molecular LUMO to initiate the proton transfer.\cite{xu2023first}
The molecular HOMO and LUMO can be hybridized with liquid water states. Thus, the hybrid states that preserve the spatial characteristics of the molecular HOMO/LUMO most closely were chosen for the initial excitation.
Figure \ref{fig:water_traj}(b) shows the results of the NEO Ehrenfest dynamics using the sc-TPB approach, and these results are compared to the case where the oHBA molecule is in the gas phase.
The NEO Ehrenfest dynamics trajectories using the TPB approach show similar results, and they can be found in Supplementary Material.
No proton transfer between the electronically excited oHBA molecule and its surrounding water molecules was observed, which is consistent with our previous study.\cite{xu2023first}

\par Previously, the ESIPT dynamics in oHBA were found to be sensitive to the configuration of the surrounding waters at the time the electronic excitation takes place using a hybrid QM/MM approach for the RT-NEO-TDDFT Ehrenfest dynamics.\cite{chow2023nuclear}
Similarly, our results for periodic QM systems indicate that the presence of surrounding water molecules qualitatively impacts the proton transfer process,
and it could even suppress ESIPT entirely, as shown by the green lines in Fig.\ref{fig:water_traj}(b).
Even though simulating only four instances is not enough to provide a comprehensive understanding of the solvation effect on the ESIPT in the oHBA molecule, these results show that solvent dynamics can play a central role in these ESIPT processes.
Additionally, we performed an analogous simulation with one of these initial molecular configurations using the conventional RT-TDDFT Ehrenfest dynamics without NEO such that all nuclei were treated classically.
In this case, the proton transfer takes place but at a much slower rate (see Figure S4 in  Supplementary Material).
This observation further emphasizes the importance of the quantum treatment of the transferring protons in proton transfer processes. 

\section{Conclusion}
\label{sec:conclusion}
\par In this work, we presented a systematic method for integrating the NEO method with periodic RT-TDDFT and Ehrenfest dynamics through the Lagrangian for multicomponent quantum-classical systems. 
The NEO Lagrangian allows us to derive the equations of motion for both quantum particles and classical nuclei on an equal footing. 
Ehrenfest dynamics with RT-NEO-TDDFT was introduced with three distinct schemes for the traveling proton basis approach, in which the centers of the quantum proton basis functions are propagated as classical degrees of freedom.
The three schemes are compared in terms of the quantum proton dynamics and the corresponding constant of motion for the dynamics. 

Building on our original periodic NEO-DFT work,\cite{xu2022nuclear} further technical details regarding the evaluation of the energy gradients and real-time propagators are provided.
The numerical implementation demonstrated that NEO Ehrenfest dynamics represents a powerful approach for investigating the coupled quantum dynamics of electrons and protons in heterogeneous environments, in which the dynamics of other nuclei are important. 
As a proof-of-principle example, we showed how the dynamics of solvating water molecules significantly impacts the photoinduced intramolecular proton transfer in an o-hydroxybenzaldehyde molecule. 
We anticipate that this new first-principles method will facilitate the investigation of a broad range of photocatalytic reactions, including PCET in extended condensed matter systems.

\section*{Supplementary Material}

\par Supplementary Material includes the validation of the implementation of RT-NEO-TDDFT Ehrenfest dynamics in the FHI-aims code, another comparison of the trajectories and conserved quantities among the TPB approaches, and the results of the simulation for the oHBA molecule solvated by liquid water using the original TPB approach and using conventional RT-TDDFT Ehrenfest dynamics.

\begin{acknowledgments}
This work is based upon work solely supported as part of the Center for Hybrid Approaches in Solar Energy to Liquid Fuels (CHASE), an Energy Innovation Hub funded by the U.S. Department of Energy, Office of Science, Office of Basic Energy Sciences under Award Number DE-SC0021173. 
We thank the Research Computing at the University of North Carolina at Chapel Hill for computing resources.
\end{acknowledgments}

\appendix

\setcounter{equation}{0}
\renewcommand{\theequation}{A\arabic{equation}}
\setcounter{subsection}{0}
\renewcommand{\thesubsection}{A\arabic{subsection}}

\section{Comment on the energy gradient for NEO-DFT}

\par
In the main text of this work, only partial spatial derivatives of energy are required when evaluating energy gradients for RT-NEO-TDDFT, where the proton density matrix response to the change of atomic coordinates is zero ($\sum_{ij}{\partial \mat{D}_{ij}^p}/{\partial \vect R_I} = 0$).
However, computing atomic forces in the ground state usually requires the evaluation of energy total derivatives with respect to atomic coordinates, particularly for proton basis function centers $R_N^p$. 
To evaluate this in the context of NEO-DFT, one combines all the terms, including the gradient of the proton density matrix in Eqs. \ref{eq:F_T}-\ref{eq:F_epc} and simplifies the sum as 
\begin{equation}
\begin{aligned}
    \sum_{ijkl,N^\prime I}{\text{d} \mat{D}_{ij}^p}/{\text{d} \vect R_N^p}\cdot & \left[ 
    \mat T^p_{ij} 
    + 2(ik|jl)\mat D^p_{kl} 
    + \int d\vect r \phi^p_i(\vect r) \phi^p_j(\vect r) V_{\text{es},N}^{p}\left( \vect R_{N}^p - \vect r \right) \right. \\
    & + \int d\vect r \phi^p_i(\vect r) \phi^p_j(\vect r) \frac{Z_I}{\left| \vect R_N^p - \vect R_I \right|}
    + \int d\vect r \phi^p_i(\vect r) \phi^p_j(\vect r) v_\text{es}^{e}\left(\vect r \right) \\
    & \left.  + \int d\vect r \phi^p_i(\vect r) \phi^p_j(\vect r) V_\text{epc}^{p}\left(\vect r \right)
    \right]
    = \sum_{ij}{\text{d} \mat{D}_{ij}^p}/{\text{d} \vect R_N^p}\cdot \mat{H}_{ij}^p.
\end{aligned}
\end{equation}
Furthermore, considering the generalized eigenvalue equation of the protonic KS system 
$$\sum_{j}\mat{H}_{ij}^p c_{j,n} = \sum_{j}\mat{S}_{ij}^p c_{j,n}\epsilon_n, $$ and the differentiation of the orthonormal equation\cite{pople1979derivative}
$$\frac{\text{d}}{{\text{d} \vect R_N^p}}(\sum_{ij} c_{i,n}^*\mat{S}_{ij}c_{j,n})=0, $$ this expression reduces to
\begin{align}
    \label{eq:F_dDdR_1}
    \sum_{ij}\frac{\text{d} \mat{D}_{ij}^p}{\text{d} \vect R_N^p}\mat{H}_{ij}^p = &
    \sum_{ij,n} f_n\left(\frac{\text{d} c_{i,n}^*}{\text{d} \vect R_N^p} \mat{H}_{ij}^p c_{j,n} + c.c.\right) \\ = &
    \label{eq:F_dDdR_2}
    \sum_{ij,n} f_n\epsilon_n \left(\frac{\text{d} c_{i,n}^*}{\text{d} \vect R_N^p}  \mat{S}_{ij}^p c_{j,n} + c.c.\right) \\ =&
    \label{eq:F_dDdR_3}
    - \sum_{ij,n} f_n\epsilon_n c^*_{i,n}c_{j,n} \frac{\partial \mat{S}_{ij}^p}{\partial \vect R_N^p},
\end{align}
where $f_n$ is the occupation number of the $n$th proton KS, $c_{i,n}$ are the coefficients of the proton KS wave functions, and $\epsilon_n$ is the $n$th KS eigenvalue. 
We note that the gradients of the proton overlap matrix ${\partial \mat{S}_{ij}^p}/{\partial \vect R_N^p}$ are negligibly small in most cases due to the localized nature of quantum protons.
Therefore, the evaluation of Eq. \ref{eq:F_dDdR_3} can be skipped in most NEO-DFT calculations. 

%

\end{document}